\documentclass[11pt]{article}

\input epsf.sty

\def\lromn#1{\uppercase\expandafter{\romannumeral#1}}

\def\lromn#1{\uppercase\expandafter{\romannumeral#1}}

\def\lromn#1{\uppercase\expandafter{\romannumeral#1}}

\begin{document}
\begin{flushright}
\today \\
%${\cal M. Y.}$
\end{flushright}

\begin{center}
\begin{Large}
\textbf{
Radiative emission of neutrino pair free of quantum electrodynamic backgrounds
}

\end{Large}
\end{center}

\vspace{1cm}
\begin{center}
\begin{large}
M. Yoshimura, 
 N. Sasao$^{\dagger}$, and M. Tanaka$^{\ddagger}$

\vspace{0.5cm}
Center of Quantum Universe, Faculty of
Science, Okayama University \\
Tsushima-naka 3-1-1 Kita-ku Okayama
700-8530 Japan

$^{\dagger}$
Research Core for Extreme Quantum World,
Okayama University \\
Tsushima-naka 3-1-1 Kita-ku Okayama
700-8530 Japan \\
$^{\ddagger}$
Department of Physics, Graduate School of Science,
Osaka University \\
Toyonaka, Osaka 560-0043 Japan

\end{large}
\end{center}

\vspace{2cm}

\begin{center}
\begin{Large}
{\bf ABSTRACT}
\end{Large}
\end{center}

A scheme of quantum electrodynamic (QED) background-free radiative emission
of neutrino pair (RENP) is proposed in order to
achieve precision determination of neutrino properties so far
not accessible.
The important point for the background rejection  is  the fact that the dispersion relation
between wave vector along propagating direction in wave guide 
(and in a photonic-crystal type fiber) and frequency
is modified by a discretized non-vanishing effective mass.
This effective mass acts as a cutoff of allowed frequencies, and one may
select the RENP photon energy region free of all
macro-coherently amplified QED processes
by choosing the cutoff larger than the mass of neutrinos.

\vspace{2cm}

Key words

neutrino mass,
relic neutrino,
Majorana particle,
macro-coherence,
photonic crystal

\newpage
\section 
{\bf Introduction}

With the advent of successful macro-coherent amplification
(more than $10^{15}$)
of QED rare processes \cite{psr observation}, namely
the macro-coherent paired super-radiance (PSR) \cite{macro-coherence intuitive},
the atomic project of neutrino mass spectroscopy \cite{renp overview}
has gained  a new stage of potentiality to explore 
important neutrino properties yet to be measured and to ultimately detect the
relic neutrino of temperature 1.9 K \cite{yst relic}.

In the present work we address the problem of quantum electrodynamic (QED) backgrounds
and propose a scheme of QED background-free RENP (Radiative Emission of Neutrino Pair).
Usual higher order QED processes, when they occur spontaneously, are not at all serious
backgrounds to macro-coherently amplified atomic neutrino pair emission (RENP)
if experiments of the neutrino
mass spectroscopy are designed with a  repetition cycle sufficiently faster than 
the decay lifetime. (Even a repetition scheme  slower than the decay rate is
conceivable if the dead time is not too large.)
The macro-coherently amplified QED  process may however become a  serious source
of backgrounds, since their rates are much larger, as is made evident below.
We shall term macro-coherently amplified QED process of order n as
McQn for brevity.
The case of $n=2$ corresponds to PSR.
Since RENP process occurs with parity change, the main backgrounds are
odd McQn.

Our proposal for the QED background rejection is to use either wave guides \cite{wave guide} or
some type of photonic crystals \cite{photonic crystal} to host a target.
A promising host is Bragg fiber consisting a hollow surrounded by two periodically
arranged dielectrics of a cylindrical shape \cite{bragg fiber}, \cite{bragg fiber 2}.
For brevity we call these hosts as host guides in the present work.
After we show below how QED backgrounds are rejected,  we shall calculate spectral rates of RENP
(stimulated single photon emission) $|e\rangle \rightarrow |g\rangle
+ \gamma_0 + \nu\bar{\nu}$ in which no background of McQ3
$|e\rangle \rightarrow |g\rangle + \gamma_0 + \gamma_1\gamma_2$ exists.
Rejection of McQ5 and so on is then automatically guaranteed.
The background-free RENP rate is, to a good approximation,
found to be a shifted (to the higher energy side)  spectrum in free space.

Parities of two states, $|e\rangle, |g\rangle$, for RENP are different.
A good example of candidate de-excitation is from the Xe excited state of
$J^P = 1^-$ of configuration $5p^5 6s$(8.437 eV) (the decay rate being $\sim$ 300 MHz) to the ground state of $0^+$ of $5p^6$.

\begin{figure*}[htbp]
 \begin{center}
 \epsfxsize=0.6\textwidth
 \centerline{\epsfbox{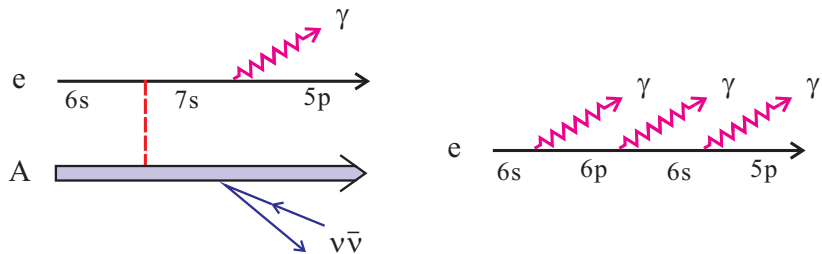}} \hspace*{\fill}
   \caption{A part of Feynman diagrams for Xe $^3P_1(8.437 {\rm eV})$ RENP in the left and
for McQ3 in the right.
Dashed red line in the left is for Coulomb excitation between nucleus A
and a valence electron.
}
   \label {diagrams}
 \end{center} 
\end{figure*}

\begin{figure*}[htbp]
 \begin{center}
 \epsfxsize=0.4\textwidth
 \centerline{\epsfbox{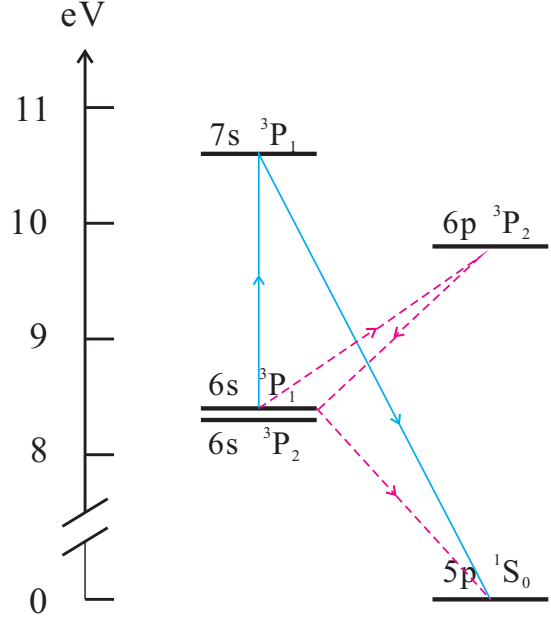}} \hspace*{\fill}
   \caption{Relevant Xe energy levels: $|e \rangle = 6s ^3P_1\,, |g \rangle = 5p ^1\!S_0\,,
|q\rangle = 7s ^3P_1$ ($6s ^3P_2$ is for the spin current contribution in Appendix).
Solid blue lines are the route for RENP, while dashed red lines are for McQ3.
}
   \label {xe levels}
 \end{center} 
\end{figure*}

We show a part of  Feynman diagrams (despite of the use of the non-relativistic
perturbation theory based on bound state electrons) in 
Fig(\ref {diagrams}).
RENP diagrams are based on the nuclear monopole contribution of \cite{nuclear monopole}.
Relevant Xe energy levels are shown in Fig(\ref {xe levels}).

The rest of this paper is organized as follows.
In Section 2 
the main background source of QED processes, McQ3, is discussed in detail and its rate
is given along with some features of background events.
Results are presented for the Xe de-excitation and the necessity of
all QED background rejection is emphasized.
In Section 3 we point out the possibility of the QED background rejection
in targets hosted in wave guides or photonic crystals, and
derive a condition for the background rejection in RENP.
The idea is based on the difference of the very nature
of freely propagating neutrino and restricted propagating 
light field in the hosted target.
The background-free RENP spectrum is approximately found to be
the shifted  spectrum in free space, the shifted
 threshold being  determined by the frequency cutoff in host guide.
A few examples of the shifted spectrum
due to the nuclear monopole contribution \cite{nuclear monopole}
are shown taking a few choices of the size  of host guides.
In the first Appendix some rudimentary facts on
propagating modes in wave guides and photonic-crystal
fibers are explained.
The second Appendix presents the  modified RENP spectrum
from $6s ^3P_2$ (8.315 eV)
due to the spin current contribution \cite{dpsty}.

Throughout this paper, we shall use the natural unit of $\hbar = c = 1$.
Conversion factors are thus
$(1 {\rm eV})^{-1} = 1.97 \times 10^{-5} {\rm cm} = (\frac{3}{1.97} \times 10^{15} {\rm Hz})^{-1}$.

\vspace{0.5cm}
\section 
 {\bf McQ3 rate}

We shall first demonstrate that without an experimental method of
background rejection McQ3 process gives an enormous background rate
when the process is macro-coherently amplified.
Thus, it becomes a necessity to invent the rejection method
when one wants to detect RENP.

%\vspace{0.5cm}
{\bf Kinematics}

For a fixed trigger frequency  $\omega_0$ (of less than $\epsilon_{eg}/2$,
with $\epsilon_{eg}$ the level spacing,
as a necessary condition for the macro-coherence),
there is a unique relation between the detected photon energy $\omega_i$ and
its emitted angle $\theta_i$:
\begin{eqnarray}
&&
\cos \theta_i = \frac{\epsilon_{eg}}{\omega_0} - 1 - \frac{\epsilon_{eg}(\epsilon_{eg} - 2 \omega_0)}
{2\omega_0 \omega_i}
\,, \hspace{0.3cm}
i = 1,2
\,.
\label {emitted direction}
\end{eqnarray}
This relation may be derived from the 4-vector relation of
$a- k_0 = -(k_1 + k_2)$ where $a=(\epsilon_{eg}, \vec{0})$
is the energy-momentum 4-vector relevant to atomic transition 
(the difference between the initial and the final states), and
$k_i, i=0, 1,2$ are those of three photons.
Note that with the macro-coherence amplification
both of the energy and the momentum are conserved
in the atomic process \cite{macro-coherence intuitive}.
The angular constraint $|\cos \theta_i| \leq 1$ gives 
\(\:
\frac{\epsilon_{eg}}{2} - \omega_0 \leq \omega_i \leq \frac{\epsilon_{eg}}{2}
\,.
\:\)

\vspace{0.5cm }
{\bf Basic $^{ 131}$Xe data:  6s$^3$P$_1$(8.437 eV) de-excitation}

The initial metastable state $|e\rangle $ is an electron-hole system consisting of a valence electron of 6s and
a hole of 5p or a filled state 5p$^5$,  their quantum numbers 
being much like those of two valence electron system.
Relevant energy levels used are
\begin{eqnarray}
&&
|e \rangle = 6s ^3P_1 (8.437 {\rm eV})
\,, \hspace{0.5cm}
|p \rangle = 6p ^3P_2 (9.821 {\rm eV})
\,, 
\\ &&
|q \rangle = |e\rangle 
\,, \hspace{0.5cm}
|g \rangle = 5p ^1S_0 (0\,  {\rm eV})
\,.
\end{eqnarray}
We have used the notation of $LS$ coupling scheme, although
other coupling scheme may be more appropriate
for Xe.
Two decay rates or A-coefficients for McQ3 rate computation are
known:
\begin{eqnarray}
&&
A(6s \rightarrow 5p) \;(  {\rm of} \; 
|q\rangle \rightarrow |g\rangle )= 2.81 \times 10^8 {\rm Hz}
\,, \hspace{0.5cm}
A(6p \rightarrow 6s) 
\;(  {\rm of} \; 
|p\rangle \rightarrow |q\rangle )= 2.24 \times 10^7 {\rm Hz}
\,.
\end{eqnarray}

\vspace{0.5cm}
{\bf McQ3 rate in free space}

We need two more atomic levels in addition to $|e \rangle, |g\rangle$, to induce McQ3:
they are denoted by $|p \rangle, |q \rangle$.
These levels are connected to each other  by
electric dipole transitions of  moments,
$d_{ep}, d_{pg}, d_{qg}$.
Quantum mechanical amplitude of McQ3 for a single route via
virtual levels $|e \rangle \rightarrow |p\rangle \rightarrow |q\rangle \rightarrow |g\rangle$ 
is in proportion to 
\begin{eqnarray}
&&
\vec{d}_{ep} \cdot \vec{E}_a \vec{d}_{pq} \cdot \vec{E}_b \vec{d}_{qg} \cdot \vec{E}_c 
\,,
\label {3 dE factors}
\end{eqnarray}
where $a,b,c$ are taken respectively from (0,1,2).
The magnitudes of dipole moments may be replaced by
A-coefficient divided by the third power of their level spacings,
hence in principle measured data can be used.

In the example of Xe de-excitation, three photon emission occurs in
the third order of QED:
\begin{eqnarray}
&&
(6s 5p^-)_{J=1} \rightarrow (6p 5p^-)_{J=2} + \gamma
\rightarrow (6s 5p^-)_{J=1}+ \gamma + \gamma \rightarrow (5p^2)_{J=0} + \gamma+
\gamma + \gamma
\,,
\end{eqnarray}
where $5p^-$ denotes a hole in $5p$ state.
The final state denoted by $5p^2$ is actually
the completely filled six $5p$ states 5p$^6$.
Three emitted photons here are taken from $\gamma_0, \gamma_1, \gamma_2$,
hence there are 3! = 6 diagramatic contributions.

\begin{figure*}[htbp]
 \begin{center}
 \epsfxsize=0.5\textwidth
 \centerline{\epsfbox{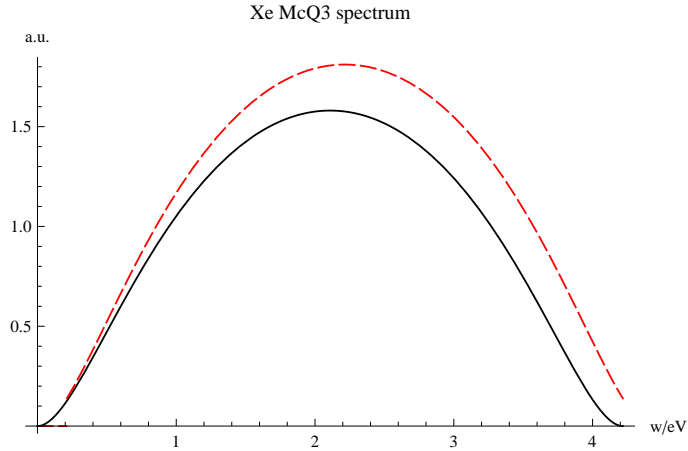}} \hspace*{\fill}
   \caption{Xe McQ3 spectrum rate in arbitrary unit.
The actual rate values  integrated over photon energies are 
$(5\,,  6) \times 10^{19}$Hz $(n/10^{20}{\rm cm}^{-3})^3 (V/{\rm cm}^3) (\eta_3(t)/10^{-3})$
for these two cases of  the trigger energy $\omega_0$:
$\omega_0 = \epsilon_{eg}/2$ in solid black and  5\% reduced value in dashed red.
}
   \label {xe mcq3 spectrum}
 \end{center} 
\end{figure*}

The differential McQ3 spectrum for a single detected photon of energy
$\omega = \omega_1$ summed over
polarizations is calculated as
\cite{identical boson effect}
\begin{eqnarray}
&&
\frac{d\Gamma_3}{d\omega} = 
\frac{3\pi^2}{2}\frac{\gamma_{pe} \gamma_{pq}\gamma_{qg}}{\epsilon_{pe}^3 \epsilon_{pq}^3  \epsilon_{qg}^3} n^3 V \eta_3(t)
\omega^2 (\epsilon_{eg} - \omega_0 - \omega)^2 F_3^2(\omega)
\,,
\label {mcq3 spectrum rate}
\\ &&
F_3(\omega) = 
\frac{1}{\epsilon_{pe} + \omega_0} 
( \frac{1}{\epsilon_{qe} +\omega_0 + \omega} + \frac{1}{\epsilon_{qg} - \omega})
+
\frac{1}{\epsilon_{pe} + \omega} 
( \frac{1}{\epsilon_{qe} +\omega_0 + \omega} + \frac{1}{\epsilon_{qg} - \omega_0})
\nonumber \\ &&
+
\frac{1}{\epsilon_{pg} - \omega_0 -\omega}(\frac{1}{\epsilon_{qg} - \omega}+ \frac{1}{\epsilon_{qg} - \omega_0})
\,.
\end{eqnarray}
The spectrum shapes are illustrated for two choices of $\omega_0$ taken close to $\epsilon_{eg}/2$;
at $\epsilon_{eg}/2$ and its 5\% reduced energy in Fig(\ref{xe mcq3 spectrum}).
Despite of a non-trivial matrix element  $\propto F_3(\omega)$,
the spectrum shape is symmetric around the point $\omega = (\epsilon_{eg} - \omega_0)/2$.
This McQ3 rate  is expected much larger than the RENP rate given below.

It is important to mention that
McQn processes, in particular, the McQ3  case
has an interesting potentiality of creating 
a quantum entangled pair of  photons under
the influence of the trigger, although this is not
the appropriate place to discuss the subject in detail

\vspace{0.5cm}
{\bf McQ3 in wave guide}

Although photonic-crystal fibers such as the Bragg fiber \cite{bragg fiber} are more promising due to
their smaller transverse sizes and other reasons,
we shall  explain our ideas using the  wave guide
because concepts are simpler there.
A good comparison of these two materials is given in \cite{photonic crystal},
in particular in its chapter 9.

As explained in Appendix,
propagation of light field in wave guide is restricted:
compared to propagation in free space,
allowed modes in wave guides have
discretized transverse wave vectors,  $\vec{k}_{\perp}$,
and their smallest cutoff.

Due to this restriction, the continuous momentum integration over 3-vector $\vec{k}$,
is replaced by a semi-discrete sum.
In the square wave guide of size $d$ these
discrete modes are given by wave vectors in two dimensions,
\begin{eqnarray}
&&
\vec{k}_{\perp} = \frac{\pi}{d} (n_1, n_2)
\,, \hspace{0.5cm}
n_i = 0, 1, 2, \cdots
\,, \hspace{0.5cm}
\sqrt{n_1^2 + n_2^2} \geq 1
\,.
\end{eqnarray}

The discrete sum over transverse momentum components
may however be approximated, , when $d\epsilon_{eg} \gg 1$, by a formula of
the continuous momentum integration,  however with the important difference
in the presence of the transverse cutoff at $k_c=\pi/d$.

\vspace{0.5cm}
\section 
 {\bf Background-free RENP rate}

%\vspace{0.5cm}
{\bf Effective cutoff in host guides}

The  dispersion relation between the energy $\omega$
and the longitudinal momentum $k_{\parallel}$ along the propagation axis  
is given by
$\omega = \sqrt{k_{\parallel}^2 + M^2}$
in host guides.
In the case of square wave guide of size $d$
the effective mass $M$ is discretized as $M = \sqrt{n_1^2 + n_2^2}\, k_c \,, k_c = \pi/d$
with $n_i$ integers, as explained in Appendix.
A similar effective mass exists in Bragg fibers along with the band gap \cite{bragg fiber}.
It is important to note that the effective mass $M$ cannot vanish.

A derivation of the condition for the QED background rejection in RENP
is given for the general case of McQn (n $\geq 3$).
We use the Lorentz invariance of $A_0^2 - A_z^2$ in the temporal and the spatial components
of 4-vector $A$, ($A_0, A_z$), where z-axis is taken along the propagation direction.
We use the energy-momentum conservation:
$a-k_0 = - k - \sum_{i=2}^{n-1} k_i$, to give a scalar invariant relation
among 2-vectors,
\begin{eqnarray}
&&
(a-k_0)^2 = (k +  \sum_{i=2}^{n-1} k_i)^2
\,.
\end{eqnarray}
The invariant left hand side (LHS) quantity may be evaluated in the laboratory
frame, to give  $\epsilon_{eg}^2 - 2 \omega_0 \epsilon_{eg} + M_0^2$,
while the right hand side (RHS) quantity is calculated in the CM system
of $n-1$ longitudinal momenta, to give
$(\omega + \sum_{i=2}^{n-1}\sqrt{k_{i,z}^2 + M_i^2} )^2$.
Hence,
\begin{eqnarray}
&&
\epsilon_{eg}^2 - 2 \omega_0 \epsilon_{eg} + M_0^2 
= (\omega + \sum_{i=2}^{n-1}\sqrt{k_{i,z}^2 + M_i^2} )^2
\,.
\label {mcqn equality}
\end{eqnarray}
For the largest $\omega_0$ value of allowed region,
one takes the smallest of RHS, namely $n-2$ photon's transverse masses set
at the fundamental mode, $M_i = k_c, i = 2\sim n-1$.
The result is expressed by the inequality,
\begin{eqnarray}
&&
\omega_0 \leq \frac{ \epsilon_{eg} }{2} - \frac{ 1}{ 2\epsilon_{eg}}
\left( (\omega + (n-2) k_c )^2 - M_0^2
\right)
\,.
\label {mcqn condition}
\end{eqnarray}
We may define the largest McQ3 trigger energy $\omega_c$
by taking $n=3, \omega=k_c$,
\begin{eqnarray}
&&
\omega_c = \frac{ \epsilon_{eg} }{2} - \frac{ 1}{ 2\epsilon_{eg}}
\left( 4 k_c^2  - M_0^2
\right)
\,.
\label {smallest trigger energy for bcgr-free renp}
\end{eqnarray}
The trigger effective mass $M_0$ is often taken to be
$k_c$ in the present work.

On the other hand,
a massive neutrino of mass $m$ satisfies the dispersion relation
of the energy and the momentum,
$E = \sqrt{p_z^2 + \vec{p}_{\perp}^2 + m^2}$.
A big difference is in transverse components:
the neutrino transverse momentum may vanish: $\vec{p}_{\perp}=0$.
This gives an allowed region for RENP;
\begin{eqnarray}
&&
\omega_0 \leq \frac{ \epsilon_{eg} }{2} - \frac{ 1}{ 2\epsilon_{eg}}
\left(  4m^2 - M_0^2
\right)
\,.
\label {renp condition}
\end{eqnarray}
RENP  thresholds of neutrino pair production of mass $m_i$ start at
\begin{eqnarray}
&&
\omega_{ii}' = \omega_{ii} + \frac{M_0^2}{2\epsilon_{eg}}
\,, \hspace{0.5cm}
\omega_{ii} =  \frac{ \epsilon_{eg} }{2} - \frac{ 2m_i^2}{\epsilon_{eg}}
\,.
\label {neutrino pair threshold}
\end{eqnarray}
This location $\omega_{ii}'$ is shifted to the higher energy side by an amount $M_0^2/(2\epsilon_{eg})$
from the free space value $\omega_{ii}$.
$M_0 = k_c$ for the trigger fundamental mode.

In both of inequalities, eq.(\ref {mcqn condition}) and eq.(\ref {renp condition}),
the special cases of equalities give the boundary curves of two allowed regions.
Thus, there exists a gap between RENP allowed regions and McQn allowed regions,
if
\begin{eqnarray}
&&
(n-1) k_c > 2 m_0
\,, \hspace{0.5cm}
n \geq 3
\,,
\end{eqnarray}
where $m_0$ is the smallest neutrino mass.
The extra $k_c$ factor here arises from the smallest
energy of $\omega$ equal to $k_c$.
In the gap region RENP is QED background-free.
The argument breaks down at $n=2$, where the usual
relation $\omega=\omega_0 = \epsilon_{eg}/2+ k_c^2/(2\epsilon_{eg})$ holds
\cite{psr background}.

\begin{figure*}[htbp]
 \begin{center}
 \epsfxsize=0.4\textwidth
 \centerline{\epsfbox{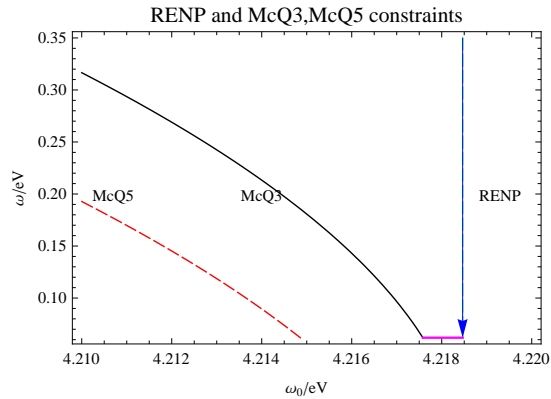}} \hspace*{\fill}
   \caption{QED background-free regions of Xe RENP in $(\omega_0, \omega)$ plane.
Two curves give upper boundaries of McQn (n=3 in solid black, =5 in dashed red) allowed regions.
Boundaries are given for the fundamental mode ($n_1^2 + n_2^2=1$) and those for excited modes (not shown here) are
all within these allowed regions.
The  background-free RENP is possible in the indicated horizontal $\omega_0-$region in red
up to the arrow point.
Assumed parameters are $\epsilon_{eg}=$8.315 eV,
and the size $d=10 \mu$m of photonic-crystal fiber.
}
   \label {ng background-free region}
 \end{center} 
\end{figure*}

Thus,
all QED backgrounds are rejected once it is rejected at the smallest
n value equal to 3.
We illustrated the background-free region in Fig(\ref{ng background-free region}).

%\vspace{0.5cm}
{\bf RENP rates}

We consider RENP from 6s$^3P_1$(8.437 eV) \cite{repetition cycle and magnetic mixing}.
We need data of 7s$^3$P$_1$(10.593 eV),
and its A-coefficient, 
$A(7s \rightarrow 5p)= 8.51 \times 10^7$s$^{-1}$, since
the Coulomb excitation bridging
between the nucleus and the valence electron requires another state of
the same quantum number $J^P$ of 6s$^3P_1$.
See  \cite{nuclear monopole} for more details.
Xe RENP rate from the nuclear pair emission,
both in free space and in host guides (assuming the
trigger field to be an allowed propagating mode), is given by
\begin{eqnarray}
&&
\Gamma_{\gamma 2\nu}(\omega ;\, t) = \Gamma_{0}F_X^2(\omega) 
I(\omega)\eta_{\omega}(t) 
\,, %\hspace{0.5cm}
\\ &&
\Gamma_0 
\sim 54 {\rm mHz} \frac{\epsilon_{eg}}{{\rm eV}} (\frac{n}{10^{21}{\rm cm}^{-3}})^3 
\frac{V}{10^2 {\rm cm}^3}
(\frac{100 {\rm MHz}}{{\rm eV}^3})^{-1}
\,.
\\ &&
I(\omega) = 
\sum_{i}\Delta_{i}(\omega)
I_{i}(\omega) \theta(\omega_{ii}-\omega)
\,,\hspace{0.5cm}
\omega_{ii} =  \frac{\epsilon_{eg}}{2} - \frac{2m_i^2}{\epsilon_{eg}}
\,,
\label{rnpe spectrum rate}
\\ &&
%\hspace*{-1cm}
I_{i}(\omega) =  
\frac{\omega^2}{3} + \frac{2m_i^2 \omega^2}{3 \epsilon_{eg}(\epsilon_{eg}-2\omega)} 
+ m_i^2 (1 + \delta_M)
\,,  \hspace{0.5cm}
%\label{rnpe spectrum rate 1}
\Delta_{i}(\omega) 
= \left( 1
 -  \frac{4 m_i^2}{\epsilon_{eg} (\epsilon_{eg} -2\omega) }
\right)^{1/2}
\,,
\label{rnpe spectrum rate 1}
\\ &&
F_X (\omega) =
\frac{ Q_w J_N ( \epsilon_{7s} - \epsilon_{6s}) }{ \epsilon_{7s} - \epsilon_{5p}}
\frac{1}{\sqrt{3\pi}}
\frac{d_{7s 5p}}{(\epsilon_{7s} - \epsilon_{6s} + \omega) 
(\epsilon_{6s}- \epsilon_{5p} - \omega)} 
\,,
\label{xe renp rate}
\end{eqnarray}
where $\delta_M = 1$ for the Majorana neutrino and $\delta_M =0$ for the Dirac neutrino \cite{renp overview}.

Both in wave guides and photonic crystals,
overlapping integrals containing mode functions need to be evaluated.
They typically take the form,
\begin{eqnarray}
&&
I = \int_0^d dx \cos \frac{n_x \pi x}{d} e^{-i P_x x}
\int_0^d dy \sin \frac{n_y \pi y}{d} e^{-i P_y y}
\,,
\end{eqnarray}
for TE modes.
Here $P_i$'s are transverse components of momentum of the neutrino pair.
Let us restrict the trigger to the fundamental TE$_{01}$ mode: $n_1=0, n_2=1$.
An elementary calculation gives
\begin{eqnarray}
&&
|I |^2 = 16 (\frac{\sin(P_x d/2)}{P_x})^2 (\frac{P_y \cos(P_y d/2) }{(\pi/d)^2 - P_y^2})^2
\,, \hspace{0.5cm}
P_{x,y} = p_{x,y} ^1 + p_{x,y} ^2
\,.
\end{eqnarray}

In the large size limit of $d \rightarrow \infty$,
the functions here are  all reduced to the delta functions,
\begin{eqnarray}
&&
|I |^2 \rightarrow 2( 2\pi)^2 d^2 
\delta(P_x) 
\left(\delta(P_y - \frac{\pi}{d}) + \delta(P_y + \frac{\pi}{d})
\right) 
\,.
\end{eqnarray}
The back to back momentum balance of transverse momenta $(\vec{p}_{\perp})_i$ of the neutrino pair
is broken by an amount, $\pm \pi \vec{n}_0/d$ according to this formula.
Thus, the spectrum rate calculated from these delta functions
coincides with the continuum limit calculation done 
as if the trigger were slightly off-axis.
Since the squared matrix element of RENP is a smooth of 
the transverse  momentum around $\vec{P}_{\perp} = 0$,
the approximation of large size limit, $d \rightarrow \infty$, is justified.

\begin{figure*}[htbp]
 \begin{center}
 \epsfxsize=0.6\textwidth
 \centerline{\epsfbox{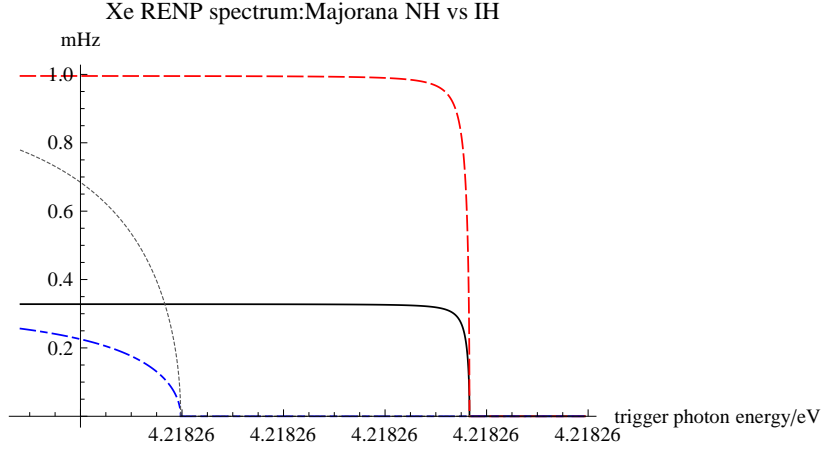}} \hspace*{\fill}
   \caption{Xe spectral rate in the background-free region for the
Majorana cases of the smallest mass, $m_0=$ 1 and 5 meV, and
NH (in solid black and in dash-dotted blue) and IH  (in  dashed red and in dotted black) .
Xe gas density of $1\times 10^{20} $cm$^{-3}$, the target
volume $1$cm$^3$, and $\eta_{\omega}  = 10^{-3}$ are assumed.
The assumed size of a host guide is 100 $\mu$m.
}
   \label {xe renp spectrum f}
 \end{center} 
\end{figure*}

As is evident in the contour map for the background-free
region, all $\omega_0-$region of $\omega_0 > \omega_c$
($\omega_c$ defined by (\ref{smallest trigger energy for bcgr-free renp})\,)
for RENP belongs to the McQn (n $\geq$ 3) background-free region.
The background-free RENP spectrum consists of
a portion of the RENP spectrum truncated at 
$\omega_c < \omega_0 < \omega_{00}' $, with
$\omega_{00}'$ determined by eq.(\ref{neutrino pair threshold})  taking the smallest neutrino mass $m_0$.
The full spectrum in the background-free region is
shown in Fig(\ref{xe renp spectrum f}).

\begin{figure*}[htbp]
 \begin{center}
 \epsfxsize=0.6\textwidth
 \centerline{\epsfbox{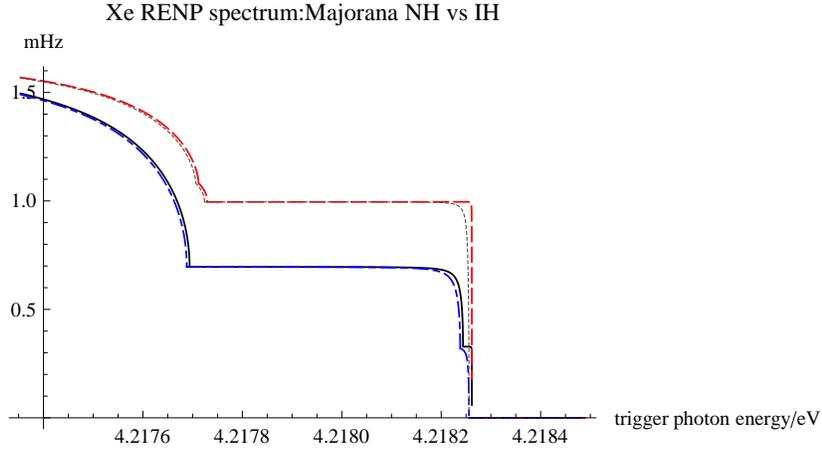}} \hspace*{\fill}
   \caption{Xe spectral rate in the background-free region for the
Majorana cases of smallest masses, 1 and 5 meV, and
NH (in solid black and in dash-dotted blue) and IH  (in  dashed red and in dotted black).
Xe gas density of $1\times 10^{20} $cm$^{-3}$, the target
volume $1$cm$^3$, and $\eta_{\omega} = 10^{-3}$.
The assumed size of a host guide is taken as 10 $\mu$m.
Three neutrino pair emission thresholds are clearly visible.
}
   \label {xe renp spectrum f2}
 \end{center} 
\end{figure*}

Finally we show in Fig(\ref{xe renp spectrum f2})
RENP spectrum in McQn $(n \geq 2)$ background-free region,
assuming $ 10 \mu$m size of photonic-crystal fiber, to show 
 all three thresholds $\omega_{ii}\,, i=1,2,3$
are measurable.
Note that the experimental resolution of $\omega_0$
is made possible by the excellent frequency resolution
of lasers, typically of order $10^{-7}$ eV or 100 MHz.

\vspace{0.5cm}
In summary,
we demonstrated that QED background-free RENP may
be realized in wave guide or in some type of photonic crystals.
With a large enough cutoff of transverse wave number
three neutrino pair emission thresholds are detectable, using
the largest nuclear monopole contribution.

\vspace{0.5cm}
We should like to thank J. Arafune for raising the QED background
problem in the correct way.

\vspace{0.5cm}
\section 
{\bf Appendix: Propagating field modes in  wave guides and photonic crystals}

Allowed  modes in rectangular wave guides are 
given by discretized transverse stationary waves.
For example, the simplest TE mode propagating in z-direction
have electric field components of the form \cite{wave guide},
\begin{eqnarray}
&&
E_x \propto \cos\frac{n_x \pi x}{a} \sin \frac{n_y \pi y}{b} e^{ik_z z} 
\,, \hspace{0.5cm}
E_y \propto \sin \frac{n_x \pi x}{a} \cos \frac{n_y \pi y}{b}e^{ik_z z} 
\,,
\end{eqnarray}
with $E_z = 0$.
$n_x, n_y$ are both integers.
With $a < b$ the lowest fundamental modes are TE$_{01}$ and TM$_{01}$
of the eigenvalue  
$-\Delta_2 = -\partial^2/\partial x^2 - \partial^2/\partial y^2 $ is $ (\pi/b)^2$.
For simplicity of notations we shall take the limit of a square wave guide, $a=b \equiv d$,
however without further consideration of degeneracy of modes.
These mode functions are orthogonal, and form
a complete set for the general expansion of propagating light waves.
The lowest excitation mode of TE$_{10}$ is  given by a non-vanishing field of
\begin{eqnarray}
&&
E_y(x,y,z; t) = 
\sqrt{\frac{\omega}{V}} 
\sin (\frac{\pi}{d }y) e^{ik_z z} e^{-i \omega t}
\,, \hspace{0.5cm}
E_x(x,y,z; t) = 0
\,, \hspace{0.5cm}
\omega= \sqrt{(\frac{ \pi}{d })^2 + k_z^2}
\end{eqnarray}
with $V = d^2 L$ the volume of the target.
The normalization of fields is determined by the condition
that the energy density of this field per unit length is $\omega/L$.

Allowed projected momentum $\vec{k}_{\perp} $($= \pi (n_x, n_y)/d$ for TE modes)  of the light field
forms on a lattice  given by discrete points
in the projected two dimensional space orthogonal to the trigger axis,
denoted in general by $\pi \vec{n}/d$.
Unless the projected momentum falls on these lattice points,
the photon emission is forbidden in  wave guide.
We may introduce an effective cutoff wave number by $k_c =\pi/d $:
transverse momenta of $|\vec{k}_{\perp} | < k_c$ are forbidden in the wave guide.
The important point for the background rejection is
that this cutoff acts as an effective mass to parallel momentum component
$k_{\parallel}$ of $\omega = \sqrt{k_{\parallel}^2 + \vec{k}_{\perp}^2} $. 

\vspace{0.5cm}
{\bf Eigenvalue problem in photonic crystals}

We shall recapitulate basic features of eigen-modes
that propagate in cylindrical fibers of photonic crystals.

First, the fundamental eigenvalue equation is set up
for the magnetic field \cite{photonic crystal},
\begin{eqnarray}
&&
\vec{H}_{\vec{k}}(\vec{r}) = e^{i\vec{k}\cdot \vec{r}} \vec{u}_{\vec{k}}(\vec{r})
\,, \hspace{0.5cm}
\vec{E}_{\vec{k}}(\vec{r}) = e^{i\vec{k}\cdot \vec{r}} 
\frac{i}{\omega(\vec{k})\epsilon(\vec{r})} (i\vec{k} + \vec{\nabla}) \times
\vec{u}_{\vec{k}}(\vec{r})
\,,
\\ &&
(i \vec{k} + \vec{\nabla}) \times \frac{1}{\epsilon(\vec{r})} (i \vec{k} + \vec{\nabla}) \times \vec{u}_{\vec{k}}(\vec{r})
= \omega^2(\vec{k}) \vec{u}_{\vec{k}}(\vec{r})
\,,
\end{eqnarray}
where $\epsilon(\vec{r}) $ is the dielectric function which may depend on spatial places.
The linear operator acting the vector function $\vec{u}_{\vec{k}}(\vec{r})$
is hermitian.

In the case of cylindrical photonic-crystal fibers, one may use the
cylindrical symmetry of the dielectric function $\epsilon(\vec{r}) $, to factor
out the cyclic coordinate dependence,
\begin{eqnarray}
&&
 \vec{u}_{\vec{k}}(\vec{r}) = e^{i k_z z + i n \varphi} \vec{h}_{k_z, n}(\rho)
\,, \hspace{0.5cm}
\rho = \sqrt{x^2+ y^2}
\,,
\end{eqnarray}
where $k_z$ is an arbitrary real number and $n$ is an integer.
We shall not write down an ordinary differential equation for
the vector function $\vec{h}_{k_z, n}(\rho)$.
This is the eigenvalue problem in one dimension (1D) due to
the cylindrical symmetry.

When the eigenvalue problem is applied to a multi-layer dielectrics of
alternating dielectric constants with a large-size air hole in the center \cite{bragg fiber},
there exist band gaps and their dispersion relation $\omega(k_z)$ 
provides the effective cutoff of order $\pi/d$, where $d$ is  the cladding periodicity,
the hole radius being taken  comparable to $d$.
Some details are shown in \cite{bragg fiber 2}.

\vspace{0.5cm}
\section 
{\bf Appendix: Spin current contribution from $^3P_2$ (8.315 eV) de-excitation}

\begin{figure*}[htbp]
 \begin{center}
 \epsfxsize=0.6\textwidth
 \centerline{\epsfbox{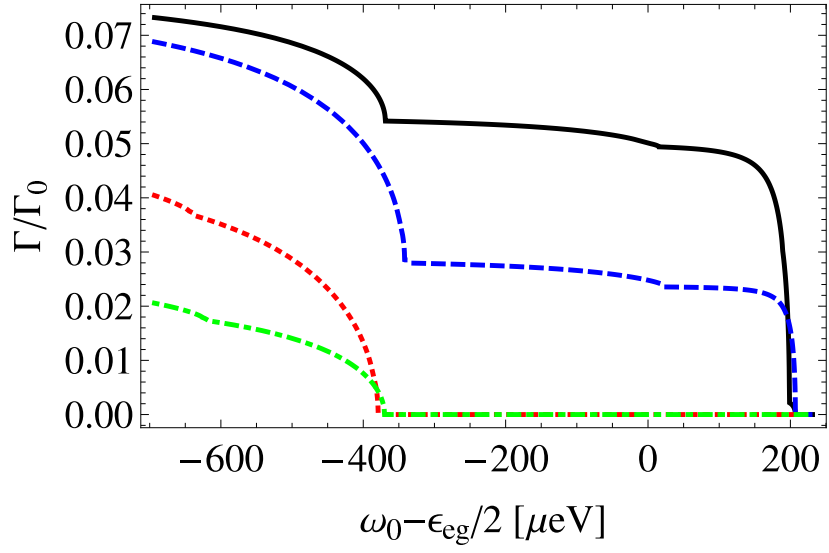}} \hspace*{\fill}\vspace*{1cm}
   \caption{Xe spectral rate in the background-free region 
due to the spin current contribution from $^3P_2$(8.315 eV) in wave guide of size
$d= 10 \mu$m.
The actual rate is derived by multiplying $\Gamma_0 \sim 9.4 \times 10^{-5}$mHz
for $n= 10^{20}{\rm cm}^{-3}, V= 1 {\rm cm}^3, \eta_{\omega} = 1$ of Xe target
(the rate scaling like $\propto n^3 V \eta_{\omega}$).
Plotted are the Majorana cases of smallest masses, 10 and 50 meV, and
NH and IH;  10 meV NH in solid black, 10 meV IH in dashed blue,
50 meV NH in dotted red, and 50 meV IH in dash-dotted green.
}
   \label {xe renp spectrum spin}
 \end{center} 
\end{figure*}

For completeness we show in
Fig(\ref {xe renp spectrum spin})
the spin current contribution using the formulas of \cite{dpsty}.
Although rates are much smaller than the nuclear monopole
contribution of preceding figures, this case has a potentiality to access the Majorana
CPV phase determination.

\begin{thebibliography}{99}
\bibitem{e-mails } yoshim@okayama-u.ac.jp

 $\dagger$ sasao@okayama-u.ac.jp 

$\ddagger$ tanaka@phys.sci.osaka-u.ac.jp

\bibitem{psr observation}
Y. Miyamoto et al,
arXiv:1406.2198v2 [physics.atom-ph] (2014)
and Prog.Theor.Exp.Phys. 113C01(2014).


\bibitem {macro-coherence intuitive}
An intuitive way of understanding the macro-coherence is as follows.
Suppose that we calculate a total rate of collective body of excited atoms given by
a formula,
\(\:
|\sum_a e^{i (\vec{k} + \vec{p}_1 + \vec{p}_2)\cdot \vec{r}_a} {\cal M}_a|^2
\,,
\:\)
where emitted plane waves at atomic site $\vec{r}_a$ are explicitly written.
If the atomic phase relaxation at different atomic sites is slow in time
and atomic matrix elements $ {\cal M}_a$ are spatially  slowly varying,
then one may factor out this part of matrix elements, to obtain
the above formula 
\(\:
= (2\pi)^3 \delta (\vec{k} + \vec{p}_1 + \vec{p}_2 )n^2V | {\cal M}|^2
\,,
\:\)
where $n$ is the number density of excited atoms in the coherent volume $V$.
An extra $n$ factor in  $\propto n^3 V$ dependence of rates
 arises from the stored field, since the process is stimulated
by the stored field.

\bibitem{renp overview}
A. Fukumi et al.,
Prog.\ Theor.\ Exp.\ Phys.\ (2012) 04D002,
and earlier references cited therein.






\bibitem{yst relic}
M. Yoshimura, N. Sasao, and M. Tanaka,
{\it Experimental method of detecting relic neutrino by atomic
de-excitation}, arXiv: 1409.3648v1 (2014).


\bibitem{wave guide}
For a general introduction on wave guides,
R.E. Collin,
{\it Foundations for Microwave Engineering},
Wiley-Interscience (1992).




\bibitem{photonic crystal}
J.D. Joannopoulos, S.G. Johnson, J.N. Winn, and R.D. Meade,
{\it Photonic Crystals}, second edition, Princeton University Press (2008).

\bibitem{bragg fiber}
P. Yeh, A. Yariv, and E. Marom,
J. Opt. Soc. Am., {\bf 68}, 1196(1978).
The band structure of the photonic-crystal fiber
is shown in \cite{bragg fiber}.

\bibitem{bragg fiber 2}
Chapter 9 of \cite{photonic crystal} on photonic-crystal fibers.
In particular, Figure 15 in this chapter is illuminating to understand
differences of field propagation in wave guides and Bragg fibers.





\bibitem{nuclear monopole} 
M. Yoshimura and N. Sasao,
Phys.\ Rev.\ D \textbf{89}, 053013 (2014).
There is a serious typo of statement immediately prior to
eq.(17) in this paper.
Canceling diagrams are the leftmost of Fig.1 and two of Fig.2
instead of the stated ones which give the result of eq.(19).
All written formulas  in the paper are  correct.
A numerical typo was further corrected.

\bibitem{dpsty}
D.N. Dinh, S. Petcov, N. Sasao, M. Tanaka, and M. Yoshimura,
{\it  Phys. Lett.}{\bf B719},154(2012).


\bibitem{identical boson effect}
Identical particle effects of bosons have been neglected in this
formula.

\bibitem{psr background}
PSR process based on two dipole transitions E1 $\times$ E1
has been discussed in
M. Yoshimura, N. Sasao, and M. Tanaka,
Phys.\ Rev.\ A \textbf{86}, 013812 (2012), which does not
contribute as RENP background when one can neglect effects of parity mixture such
as a stray field in Xe de-excitation.
Other types of PSR arising, for example, from M1 $\times$ E1 type  involving parity change,
may contribute as a background in Xe RENP.
This background can be rejected if one chooses $\omega_0$ of RENP 
larger than $\epsilon_{eg}/2$, which is possible by the up-shifted neutrino pair threshold.

\bibitem{repetition cycle and magnetic mixing}
The scheme we consider here for RENP  relies on a fast repetition cycle of
back to the ground state in order not to lose RENP events in  dead time.
This is because the development of macro-coherence is destroyed at the
phase relaxation time $T_2$, which is typically 1$\sim$ 10 ns in dense targets.
A typical repetition cycle we may think of is of order 1$\sim$ 100 kHz, to be compared with
the E1 transition rate of order 10 ns, giving an effective RENP time fraction of order $10^{-5} \sim 10^{-3}$.
The effective RENP rate is reduced by this factor from those of rates in our figures.

Another scheme is to use as the initial state $^3P_2 (8.315 {\rm eV})$ with a magnetic mixing 
of the state $^3P_1 (8.437 {\rm eV})$ (required from the selection rule of $|\Delta J| = 1$ for RENP).
The magnetic mixing amplitude $\delta_m$ is of order,
\(\:
\delta_m = \mu_B B/\Delta \sim 5 \times 10^{-5} (B/{\rm T} ) ({\rm eV}/\Delta)
\,, 
\:\)
with $\Delta$ the energy difference between two mixed states ($0.12$eV in the Xe case, giving 
 $\delta_m $ of order $ 5\times 10^{-2}$ for 10 T magnetic field).
RENP rates are reduced by $\delta_m^2$.




\end{thebibliography}
\end{document}